\documentclass[aps,pre,floatfix,nofootinbib,twocolumn,superscriptaddress,amsmath,amssymb]{revtex4}
\usepackage{amsmath} \usepackage{epsfig} \usepackage{bm}

\newcommand{\actionE}{\mathcal{E}}

\begin{document}

\title{Spacetime thermodynamics and subsystem observables in a
kinetically constrained model of glassy materials}

\author{Robert L. Jack} 

\affiliation{Rudolf Peierls Centre for
Theoretical Physics, University of Oxford, 1 Keble Road, Oxford, OX1
3NP, UK}
\affiliation{Department of Chemistry, University of California,
Berkeley, CA 94720-1460}

\author{Juan P. Garrahan}

\affiliation{School of Physics and Astronomy, University of
Nottingham, Nottingham, NG7 2RD, UK}

\author{David Chandler}

\affiliation{Department of Chemistry, University of California,
Berkeley, CA 94720-1460}

\begin{abstract}
In a recent article [M. Merolle et al., Proc. Natl. Acad. Sci. USA 102,
10837 (2005)], it was argued that dynamic heterogeneity in $d$-dimensional
glass formers is a manifestation of an order-disorder phenomenon in the
$d+1$ dimensions of space-time.  By considering a dynamical
analogue of the free energy, evidence was found for phase coexistence
between active and inactive regions of space-time,
and it was suggested that this phenomenon underlies the
glass transition.  Here we develop these ideas further by
investigating in detail the one-dimensional Fredrickson-Andersen (FA)
model in which the active and inactive phases originate in
the reducibility of the dynamics.  We illustrate the phase coexistence
by considering the distributions of mesoscopic space-time observables.
We show how the analogy with 
phase coexistence can be strengthened by breaking microscopic 
reversibility in the FA model, leading to a non-equilibrium 
theory in the directed percolation universality class. 
\end{abstract}

\maketitle

\section{Introduction}

\begin{figure}[t]
\begin{centering}
\includegraphics[width=8.5cm]{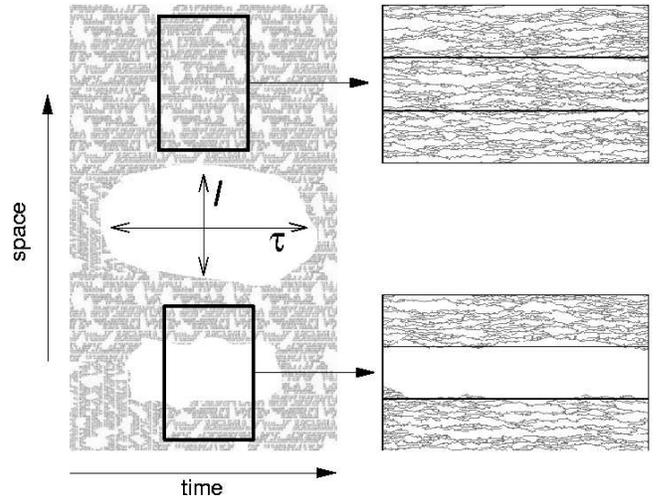}
\caption{Illustration of a trajectory in a facilitated model with
  space-time ``bubbles'' of the inactive state.  The boxes illustrate
  finite observation space-time windows: the top one corresponds to a
  typical region; the bottom one is a rare collective fluctuation of
  size much larger than those typical of the active state.  On the
  right are trajectories from the one-spin facilitated one-dimensional
  Fredrickson-Andersen model \cite{FredricksonA84}, at $T=1$ for
  observation windows of $L=180$ and $t_{\rm obs} = 320$ (smaller
  observation windows of $L=60$ are also outlined). }
\label{fig:bubble_sketch}
\end{centering}
\end{figure}

The hallmarks of the glass transition \cite{ReviewsGT} are a sudden
increase of relaxation times, and dynamic heterogeneity \cite{ReviewsDH}.  
A recent paper \cite{MerolleGC05} described
the idea that these phenomena are manifestations of 
phase coexistence in space-time. 
The two competing phases are an active state,
containing many relaxation events, and an
inactive one, where these events are scarce.
The purpose of the current paper is to develop these ideas by 
building on the results of Ref.~\cite{MerolleGC05}.

In the active phase of the dynamics, the existence of
the inactive phase 
can be inferred by measuring the distribution of any
observable that measures the quantity
of relaxation events within a finite space-time window
\cite{MerolleGC05}.
Because of the coexisting inactive phase,
these distributions display non-Gaussian tails.
This behaviour is analogous to the situation
in standard phase transitions: the distribution of cavity sizes
in a fluid near liquid vapour coexistence 
has non-Gaussian tails 
\cite{HC00}, as does the distribution of box magnetization
for the Ising model in the vicinity of its phase transition
\cite{CFH04}. 

The concept of space-time thermodynamics that we use
\cite{MerolleGC05} is illustrated in Fig.~\ref{fig:bubble_sketch}.
It employs a simple picture of glassy dynamics at low temperatures, in which
diffusing excitations coalesce and branch \cite{GarrahanC02}.  This dynamics is
reducible \cite{RitortS03}. That is, there are two steady states, an active one
with a finite density of excitations, and
an inactive one with strictly zero density.  
Working in the active state, the dominant
fluctuations on large length scales
are space-time regions in which there are no
excitations, as illustrated in Fig. \ref{fig:bubble_sketch}.  We refer
to these regions as ``bubbles'' of the inactive phase. 

We characterise the bubbles
by the probability density function
$P_\mathrm{bubble}(l,\tau)$ for their spatial and temporal extents
(denoted $l$ and $\tau$, respectively).  To the extent 
that these rare fluctuations are dynamical analogues of those near 
phase coexistence, and for bubbles which are large compared
with the bulk correlation length of the active state, we can write:
\begin{equation}
P_\mathrm{bubble}(l,\tau) \propto \exp\left\{-\left[ \sigma_1 l + \sigma_2
\tau + \mu l\tau + \Lambda (l-v\tau)^2\right]\right\} , 
\label{equ:bubble_dist}
\end{equation}
where $\sigma_{1}$ and $\sigma_2$ are, in effect, surface tensions in the spatial and
temporal directions respectively, $\mu$ is the difference between the bulk free energy 
densities of
active and inactive phases, and $\Lambda$ controls the aspect ratio of
the bubbles, in conjunction with the velocity parameter $v$.  This distribution of (\ref{equ:bubble_dist})
can be investigated by considering
the probability distribution functions for observables that are
averaged over finite regions of space-time, such as the boxes in 
Fig.\ \ref{fig:bubble_sketch}~\cite{MerolleGC05}.  
Large rare 
bubbles dominate the tails of these distributions.

The terminology ``free energy'', as used in the previous 
paragraph, refers to space-time; the relevant ensemble is
a the space of histories (or trajectories) 
of the system, which is observed for a specified 
period of time, $t_\mathrm{obs}$.  
This concept of a free energy is central 
to Ref. \cite{MerolleGC05}, and it is found also in the thermodynamic formulation of 
dynamics due to Ruelle \cite{RuelleBook,Lecomte06}.  In particular, for a $d$-dimensional 
system, one considers statistical mechanics in the corresponding 
$(d+1)$-dimensional space. The extra dimension is that of time, so the resulting
space is necessarily anisotropic. 

Thus, to appreciate the analogy with phase coexistence for $d=1$,
consider an anisotropic two-dimensional Ising model below its critical 
temperature, with a small field stabilising the spin-up phase. The spins 
are predominately up, but there will be 
a distribution of spin-down domains with various sizes. In the case 
where the interactions are different in the $x$ and $y$ directions,
the distribution of sizes for domains of the spin-down phase is
of the form
\begin{equation}
P_\mathrm{down}(x,y) \propto \exp\left\{-\left[ \sigma_x x + \sigma_y
y + \mu' xy + \Lambda' (x-v'y)^2\right]\right\},
\label{equ:ising_dist}
\end{equation}
where $x$ and $y$ are the extents of the bubbles in the $x$ and $y$
directions; $\sigma_x$ and $\sigma_y$ are surface tensions; $\mu'$ is the
free energy difference between up and down phases (proportional to
the small field); and $\Lambda'$ and $v'$ set the shape of the bubbles.

\begin{figure}
\epsfig{file=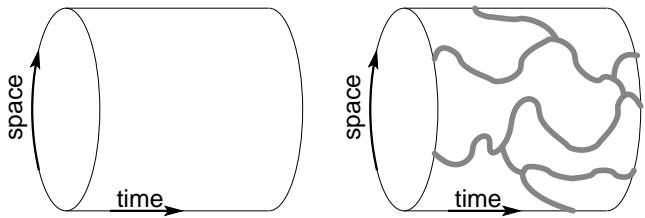,width=8.5cm}
\caption{Sketch illustrating the effect of the choice of
initial condition in a model of diffusing excitations that branch
and coalesce. An initial state with
no excitations (left) persists
throughout the observation time. All other initial conditions
result in the system exploring the active steady state (right).}
\label{fig:bc}
\end{figure}

As phase coexistence is approached in the Ising system, the free energy difference $\mu'$ 
approaches zero. In the dynamical system, 
the corresponding limit is $\mu\to0$, in which
case active and inactive states coexist.
        A central
        result of this article is that
        $\mu=0$ for systems in which the
        dynamics is reducible into active and inactive
        partitions.  Within the thermodynamic formalism of Ruelle
        the free energy difference $\mu$ is the difference in the so-called
        `topological pressure' \cite{RuelleBook,Lecomte06} 
        between the active and inactive
        phases of the dynamics.


Since the phases are at coexistence in these models, 
a boundary
condition suffices to make the system choose one
phase. For example, consider a system with periodic spatial boundary conditions, as
sketched in Fig.\ \ref{fig:bc}. 
If we choose
our initial condition to be completely inactive, 
then this 
state persists forever; all other initial conditions lead to the active state. 

In this article, we use the 
the Fredrickson-Andersen (FA) facilitated spin model
\cite{FredricksonA84} as an illustrative system.
In the following section, we
probe the distribution of (\ref{equ:bubble_dist})
by 
analysing the distributions of
space-time observables, such as the total amount of dynamical activity
inside a given spatial
region, integrated over a finite observation time.  
The forms of these distributions also imply that 
$\mu=0$, in accordance with our analytic arguments.

We also compare the FA model with two other systems.
In Ref.~\cite{JMS06}, a model with appearing and annihilating excitations
(the so-called AA model) was shown to  
have the same two point
correlations as the FA model (up to a multiplicative factor). However,
we show that this model has mesoscopic fluctuations that are very 
different from those of the FA model.
Conversely, breaking microscopic reversibility in the FA model moves
moves the system into the directed percolation (DP) universality 
class~\cite{Hinrichsen00}.
In that case, the the two point functions and critical scaling change 
qualitatively. However, the active and inactive states still coexist,
so the fluctuations are still well-described by the phase coexistence
analogy.

The paper is organized as follows.  In section~\ref{sec:model} we
define the model, discuss its trajectories, and introduce the
observables of interest.  We also show that reducibility of the
dynamics is essential to see the phase coexistence effect.  In
section~\ref{sec:action} we formalise our discussion of space-time
thermodynamics by considering the distribution of the dynamical action
that is analogous to the thermodynamic free energy.  We consider the
effect of breaking microscopic reversibility in section~\ref{sec:dp}.

\section{Models, trajectories and observables}
\label{sec:model}

Kinetically constrained models (see Ref.\ \cite{RitortS03} for a
comprehensive review) are defined in such a way that their non-trivial
behaviour is purely dynamical in origin.  Thus, they are the natural
framework for investigating dynamical heterogeneity and its
consequences
\cite{Harrowell93,GarrahanC02,GarrahanC03,JungGC04,ToninelliWBBB05,BBL05}.

A very simple kinetically constrained model is the single-spin
facilitated FA model \cite{FredricksonA84,RitortS03} in dimension
$d=1$.  It is defined for a chain of $N$ binary variables $n_i \in
\{0,1\}$, with trivial Hamiltonian $H= J \sum_i n_i$.  The model
evolves under the following Monte-Carlo (MC) dynamics.  At each
MC iteration a randomly chosen site $i$ changes state according to:
\begin{eqnarray}
n_i= 0 & \rightarrow & n_i=1 \qquad \hbox{probability } f_i e^{-\beta} ,
\nonumber
\\ n_i=1 & \rightarrow & n_i=0 \qquad \hbox{probability } f_i ,
\nonumber
\end{eqnarray}
where we have set $J=1$ and $\beta \equiv T^{-1}$.  The non-trivial
part of the dynamics is due to the facilitation function
\begin{equation}
f_i \equiv n_{i+1} + n_{i-1} - n_{i+1} n_{i-1}.
\label{equ:def_f} 
\end{equation}
That is, a spin flip on site $i$ can take place only if at least one if
its nearest neighbours is in the excited state.  Since $f_i$
does not depend on $n_i$ then the time evolution of the model obeys
detailed balance with respect to $H$ at temperature $\beta^{-1}$.  The
mean density of up spins in the active state is
\begin{equation}
c \equiv (1+e^\beta)^{-1} .
\end{equation}
The unit of time is an MC sweep ($N$ attempted spin flips).

In the following, we consider an FA model in which $N$ is to be taken
to infinity in the thermodynamic limit.  We will take a subsystem of
this model, containing $L+2$ spins, $\{ n_0, \dots, n_{L+1} \}$. The
length $L$ is to remain finite: it is a mesoscopic quantity.  A
trajectory for the subsystem specifies the state of the $(L+2)$ spins
at $N t_\mathrm{obs}$ MC steps. The trajectory is defined within a space-time
observation box of size $L t_{\rm obs}$.  We consider observables
such as the density of excitations within the box, for a given trajectory:
\begin{equation}
m_\mathrm{traj} = (LNt_{\rm obs})^{-1}
\sum_{i=1}^L \sum_{\tau=1}^{Nt_{\rm obs}} n_{i,\tau}
,
\label{equ:def_mag}
\end{equation}
where $n_{i\tau}$ is the state of the $i$th spin in the $\tau$th state
of the trajectory. Note that the boundaries $n_0$ and $n_{L+1}$ do
not appear in the sum. 

The quantity $m_\mathrm{traj}$ is analogous to the net magnetization of 
a $(d+1)$-dimensional Ising model, and thus we often refer to it as the 
``magnetization''.  However, its physical meaning is that of mobility 
or activity, i.e., the net number of cells in the observed space-time 
volume that exhibit an appreciable degree of molecular motion. 
This meaning is that of Garrahan and Chandler's coarse grained view of 
structural glasses \cite{GarrahanC02,GarrahanC03}.  
In an atomistic treatment, a 
corresponding quantity can be expressed in terms of
the observable $\hat{F}_j(\bm{k},\delta t,t) =
\cos\{
\bm{k}\cdot[\hat{\bm{r}}_j(t+\delta t)-
\hat{\bm{r}}_j(t)]\} $ where 
$\hat{\bm{r}}_j(t)$ is the position of the $j$th particle
at time $t$, and $\bm{k}$ is a microscopic wave vector
(on the order of the inverse particle size). 
If $\hat{F}_j(\bm{k},\delta t,t)$ is small, then particle $j$ 
experienced a relaxation event between times $t$ and $t+\delta t$.
For a subsystem of particles, $S$, we define the density of relaxation
events, or net mobility, in space and time to be
\begin{equation}
\hat{K}_\mathrm{traj}(\bm{k},\delta t) 
= (t_\mathrm{obs} N_S)^{-1} \sum_{j\in S} 
\int_0^{t_\mathrm{obs}} \!\mathrm{d}t\, [1-\hat{F}_j(\bm{k},\delta t,t)]
\end{equation}
where $N_S$ is the number of particles in the subsystem $S$. 
The quantity $\hat{K}_\mathrm{traj}(\bm{k},\delta t)$ 
is parametrised by microscopic
length and time scales ($|\bm{k}|^{-1}$ and $\delta t$), 
and is summed over a mesoscopic region
of space-time. The parameters $|\bm{k}|^{-1}$ and $\delta t$ 
coincide with coarse graining lengths and time, 
respectively.

The fluctuations in $\hat{K}_\mathrm{traj}(\bm{k},\delta t)$ 
are related to four point observables
\cite{ToninelliWBBB05}; the distribution of 
$\hat{K}_\mathrm{traj}(\bm{k},\delta t)$ is also
related to the distributions of correlation functions measured
in \cite{Cipelletti05,Chamon04}. Results for distributions
of the density of relaxation events in atomistic glass-formers
show non-Gaussian tails similar to those that we present 
here \cite{Lutz_kinks}. 
However, the purpose of this article is to discuss generic properties
of dynamically heterogeneous systems, so we restrict ourselves
to very simple models.

\subsection{Distribution of the event densities}

We consider the distribution of the trajectory activity, or
box magnetisation $m_\mathrm{traj}$:
\begin{equation}
P(m) = \sum_{\mathrm{traj}} P_\mathrm{traj} \delta(m-m_\mathrm{traj})
.
\end{equation}
The sum
is over all possible trajectories of the system, with their associated
probabilities $P_\mathrm{traj}$.  In the limit of $L\to\infty$ or
$t_{\rm obs}\to\infty$, the central limit theorem states that $P(m)$
will be sharply peaked around $m=c$, with a variance that scales as
$(Lt_{\rm obs})^{-1}$. The observation of Ref.\ \cite{MerolleGC05} was
that while this is the case in the limit of large $(Lt_{\rm obs})$,
the distribution of $m$ remains non-trivial even for $L \gg \xi$ and
$t_{\rm obs} \gg \tau$, where $\xi \approx c^{-1}$ and $\tau \approx
c^{-3}$ are the correlation length and time associated with the zero
temperature dynamical fixed point of this model.

\begin{figure}[t]
\begin{centering}
\includegraphics[width=8.5cm]{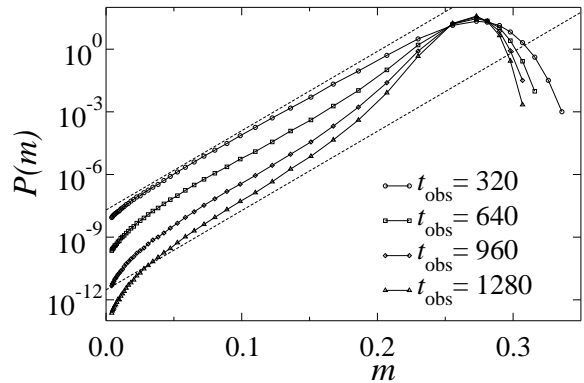}
\caption{Distribution of trajectory magnetization $P(m)$ at $\beta=1$,
$L=60$ and various observation times. We use $N=180$ which is large
enough so that $P(m)$ does not depend on $N$.  The exponential tails
of $P(m)$ all have similar gradients: the dotted lines are $P(m) \propto
\exp(\sigma m)$ with $\sigma=87$. }
\label{fig:pm_simple}
\end{centering}
\end{figure}

We show various $P(m)$ in Fig.\ \ref{fig:pm_simple}.  The peak of
$P(m)$ is Gaussian, and this corresponds to trajectories like the one
on the top-right of Fig.\ \ref{fig:bubble_sketch}.  The tails of
$P(m)$ are exponential for small $m$.  This data was obtained using a
combination of transition path sampling (TPS) \cite{BCD+02} and
umbrella sampling \cite{FrenkelSmit}. For details see the
Appendix. The trajectories in the tail are dominated by rare large
regions which have no excitations, like the the one on the
bottom-right of Fig.\ \ref{fig:bubble_sketch}.

To explain the presence of the exponential tail,
suppose that we have a trajectory with an initial condition
containing a large empty region of size $x$, and that this empty
region persists throughout $t_{\rm obs}$, leading to a box
magnetisation that is small.  Now consider a trajectory whose initial
condition has an empty region of size $x+(\Delta x)$, but is otherwise
identical to the original one.  Then the probabilities of the two
trajectories are in the ratio $(1-c)^{\Delta x}$, and their
magnetisations differ by approximately $c(\Delta x)/L$.  Thus we
predict
\begin{equation}
P(m) \sim (1-c)^{mL/c} = \exp[ (mL/c) \ln(1-c) ] .
\label{equ:pm_naive}
\end{equation}
The significance of this result is that the right hand
side is independent of
$t_{\rm obs}$: increasing the observation time changes the factor
multiplying the tail in $P(m)$, but the gradient of the tail remains constant.
Fig.\ ~\ref{fig:pm_simple} shows that the gradient is indeed independent
of $t_{\rm obs}$.  However, the quantitative prediction for its value
is accurate only to within 20-30\%. This inaccuracy arises because
the large inactive region may extend beyond
the edge of the box. 
Thus, adding extra down sites to the initial condition may
decrease the magnetisation by an amount less than $(c/L)$.

\begin{figure}[t]
\begin{centering}
\includegraphics[width=8.5cm]{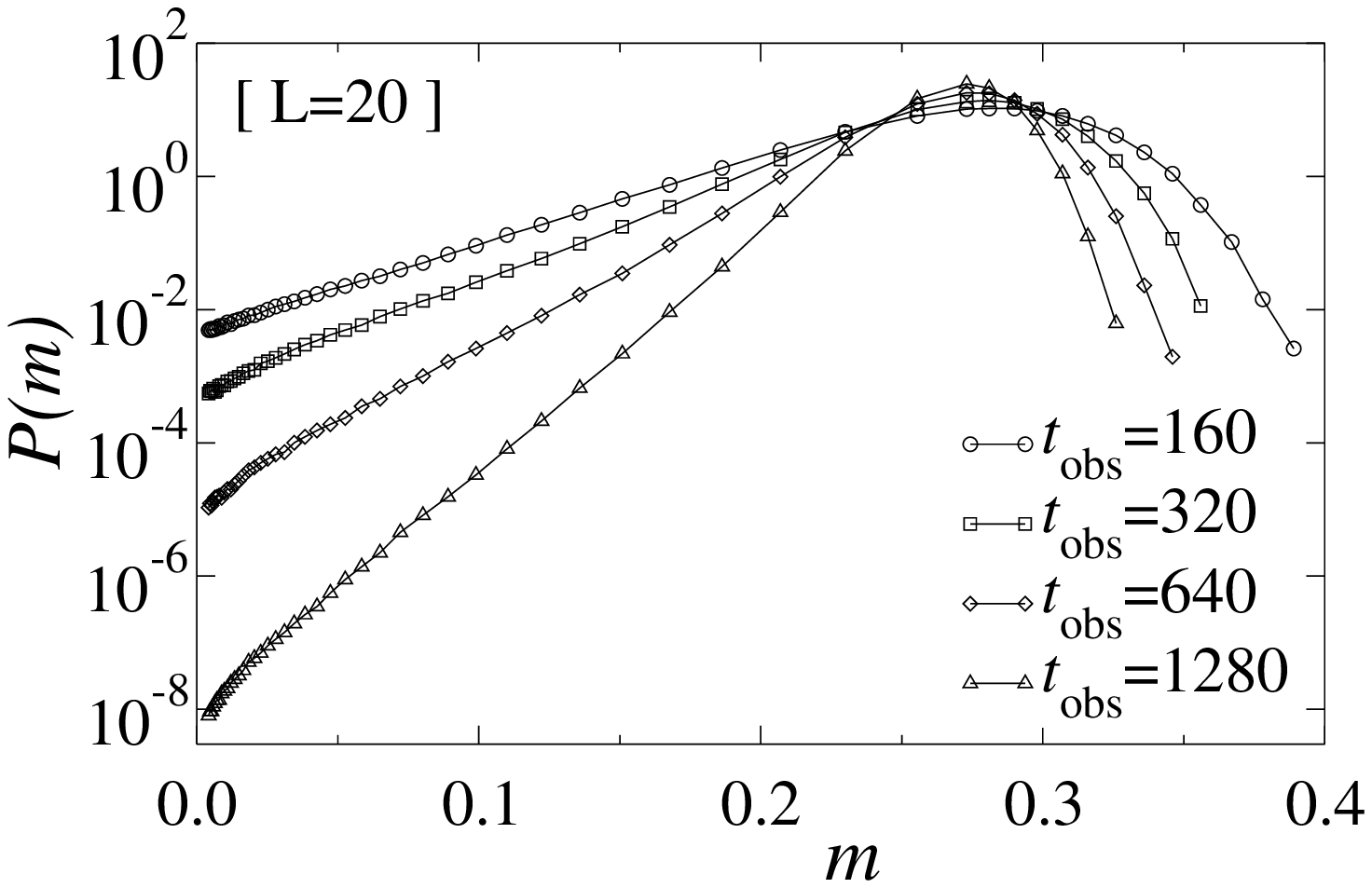}\par
\includegraphics[width=8.5cm]{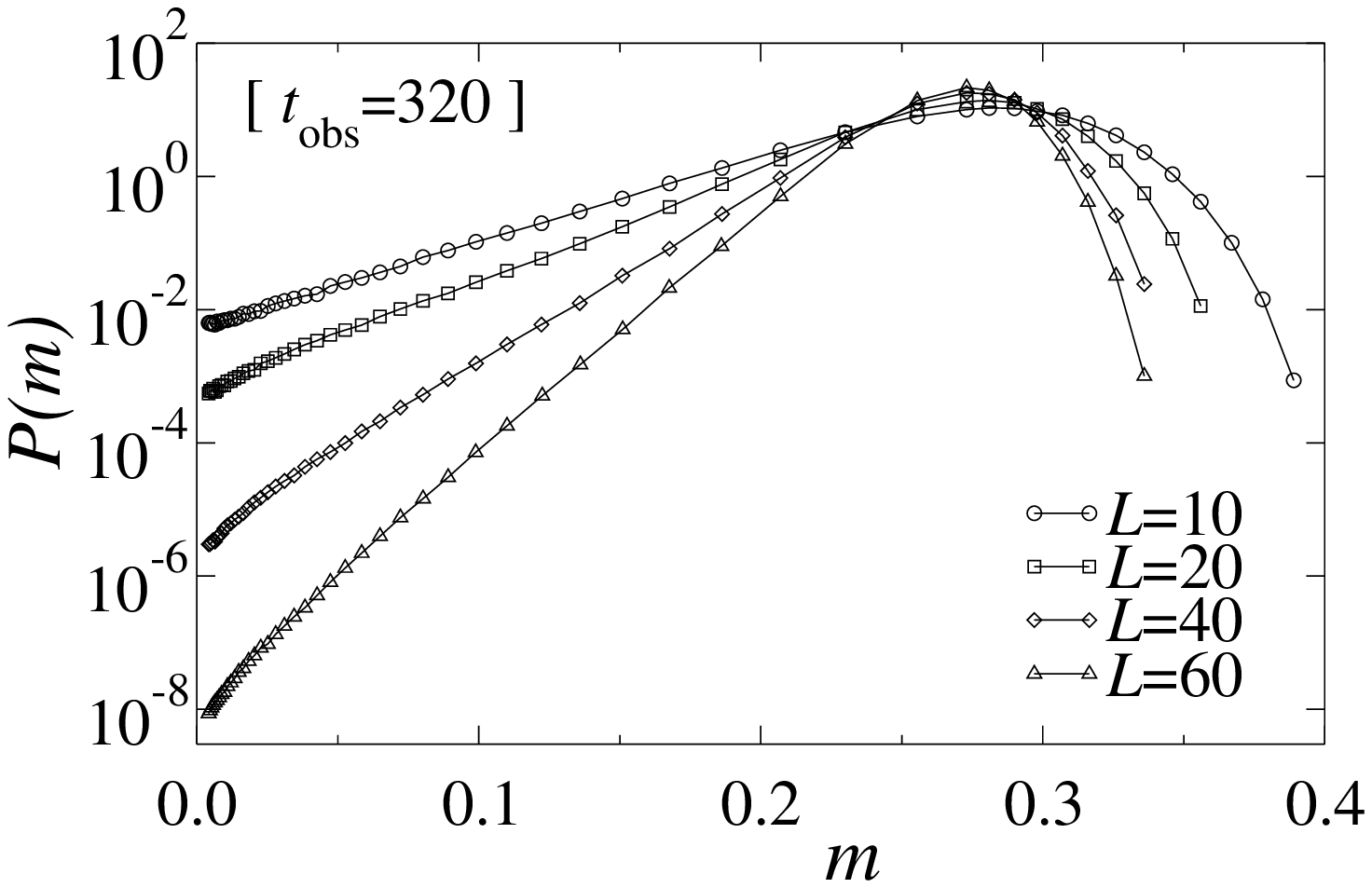}\par
\includegraphics[width=4.4cm,height=3cm]{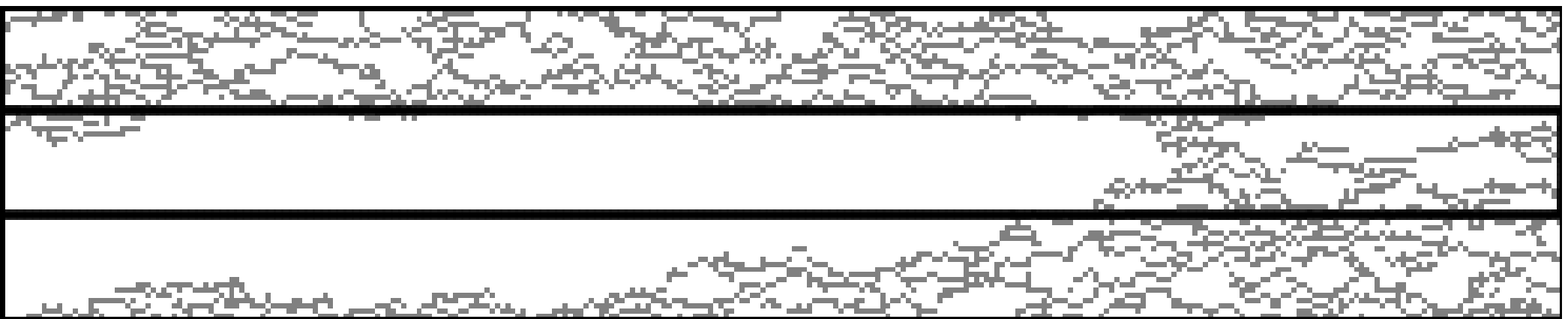}
\caption{We show $P(m)$ at $\beta=1$ for varying $L$ and $t_{\rm
obs}$.  We use $N=\max(120,3L)$ which is large enough that the results
do not depend on $N$.  (Top) Increasing observation time at
at fixed $L=20$.  (Middle) Increasing box size $L$
at $t_{\rm obs}=320$.  As $L$ or $t_{\rm obs}$ is
increased, we move from a regime in which the tail gradient is
independent of the increasing parameter to a regime in which the
gradient is proportional to that parameter.  (Bottom) We show
a typical trajectory for large $t_\mathrm{obs}$ and $L=20$ where
the observation box is outlined: the size of the total spatial region 
shown is $N=60$. For large $L$ the
trajectories are of the form shown in Fig.\ \ref{fig:bubble_sketch}.
}
\label{fig:pm_l20}
\end{centering}
\end{figure}

In Fig.\ \ref{fig:pm_simple}, the exponential tail of $P(m)$ depends
only on $L$ and not on $t_{\rm obs}$, but this is a result of the
shape of the observation box: the behaviour of $P(m)$ is symmetric
with respect to $L$ and $t_{\rm obs}$.  In Fig.\ \ref{fig:pm_l20} we
show how $P(m)$ changes as we change the aspect ratio of the
observation box.  There is a crossover at $L \sim vt_{\rm obs}$, where
$v \simeq 0.06$ at $\beta=1$. For $L>vt_{\rm obs}$ the gradient of the
exponential tail is independent of $t_{\rm obs}$ as described above,
but for $L<vt_{\rm obs}$ then the argument must be modified. Typical
trajectories at small $m$ change their form to that shown in Fig.\
\ref{fig:pm_l20} (right). In that case, increasing the width of the
bubble does not change $m$; we must instead increase the time for
which the inactive state persists. Repeating the argument leads to a
situation in which the gradient of the exponential tail is
proportional to $t_{\rm obs}$ and independent of $L$, see Fig.\
\ref{fig:pm_l20}.  In other words, the gradient of the exponential
tail is proportional to $\max(L,vt_{\rm obs})$.  The spatial and
temporal extent of the box enter on equal footing.  The data of Fig.\
\ref{fig:pm_l20} is consistent with the crossover from
large $L$ to large $t_{\rm obs}$ occuring
when $t_{\rm obs}$ is of the order of the first passage time across a
bubble of size $L$. This time scales as
$
(L/v) \sim (L/c^2)
$
since the spreading velocity \cite{Hinrichsen00}
in the FA model is $v\sim(\xi/\tau)\sim c^2$.

\begin{figure}[t]
\begin{centering}
\includegraphics[width=8.5cm]{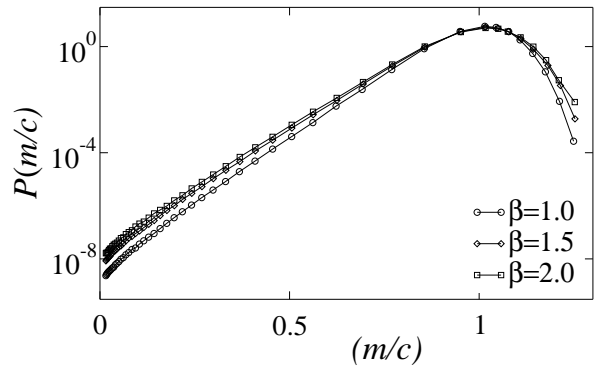}
\end{centering}
\caption{Data showing (approximate) scaling of $P(m)$ in the FA model
at various temperatures, scaled according to (\ref{equ:temp_scaling}).
We plot $P(m/c)$: the box sizes are $L=(60,88,135)$; the observation
times are $t_{\rm obs}=(320,1280,5330)$; and we use $N=3L$.  These
temperatures are not very small, so there are subleading corrections
to scaling, but there is no qualitative change to the scaled
distribution on lowering the temperature.  Further, the computational
time required at $\beta=2$ is quite significant, so we cannot rule out
small systematic errors arising from non-convergence of our TPS
procedure (see appendix).}
\label{fig:fa_scaling}
\end{figure}

Consistent with the known scaling of the FA model in one
dimension~\cite{RitortS03}, we find that the temperature
dependent of $P(m)$ can be accounted for by using rescaled
variables. The function
\begin{equation}
P(m/c; cL,c^3 t_{\rm obs})
\label{equ:temp_scaling}
\end{equation}
depends only weakly on $c$, as shown in Fig.\ \ref{fig:fa_scaling}.  
While subleading corrections
to the scaling do appear to be significant, there is no qualitative
change on lowering the temperature.  This justifies our working 
at inverse temperatures around
$\beta=1$, where the data is easier to obtain than in the
more glassy regime of the FA model ($\beta>1$).

\begin{figure}[t]
\includegraphics[width=8.5cm]{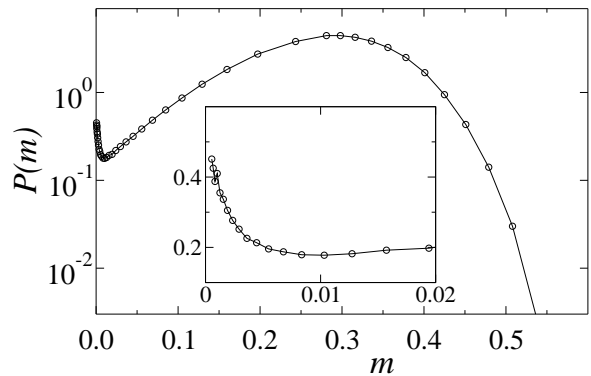}
\caption{Plot of $P(m)$ at $\beta=1$, $L=2$, $t_{\rm obs}=160$,
showing secondary maximum at small $m$. (Inset) Enlargement 
of the secondary
peak, shown in a linear scale for $P(m)$.}
\label{fig:pmfa_small}
\end{figure}

As a final observation in this subsection, we note that the behaviour
of $P(m)$ for small observation boxes, $cL < 1$ or $c^3t_{\rm obs} <
1$, is different from that for large boxes.  This is shown in Fig.\
\ref{fig:pmfa_small}.  The distribution acquires a secondary peak at
$m\to0$: there is a finite probability of observing $m=0$, so that
$P(m)$ diverges as $m \to 0$.  We will return to this feature below.

\subsection{Comparison with a model of appearing and annihilating
excitations (AA model)}

In the previous subsection, the non-trivial structure in $P(m)$ occurs
in the tails of the distribution. It is associated with rare
collective fluctuations in the system. We will
argue below that these rare fluctuations do contain
significant information about the model.  We first 
compare the magnetisation distributions of the FA
model and the AA model of \cite{JMS06}. The AA model has the same
two-point correlation functions as the FA model
in equilibrium, but other properties of the two systems are qualitatively 
different.

We define the AA model by specifying local moves for a chain
of binary variables:
\begin{eqnarray}
(n_i,n_{i+1})=(1,0) &\leftrightarrow & (n_i,n_{i+1})=(0,1),
\qquad \hbox{rate } \gamma e^{-\beta'}
\nonumber\\
(n_i,n_{i+1})=(1,1) &\rightarrow & (n_i,n_{i+1})=(0,0), 
\qquad \hbox{rate } \gamma 
\nonumber\\
(n_i,n_{i+1})=(0,0) &\rightarrow & (n_i,n_{i+1})=(1,1),
\qquad \hbox{rate } \gamma e^{-2\beta'}
\nonumber
\end{eqnarray}
where $\gamma=2/(1-e^{-\beta'})^2$ is an arbitrary temperature
dependent factor that rescales the time, chosen for later
convenience. These rates again respect detailed balance with respect
to $H=\sum_i n_i$, and the inverse temperature is $\beta'$.

Correlation functions in the steady state of the AA model at a given
temperature are related to those in the FA model at a different
temperature \cite{JMS06} (strictly this holds when the facilitation
function is proportional to the number of up neighbours, not when it
is equal to zero or one as here, but there is no qualitative change to
the physics).  The relationship between correlators depends only on
simple multiplicative factors, for example \cite{JMS06}
\begin{eqnarray}
\lefteqn{
\langle n_{it} n_{jt'} \rangle_{\mathrm{FA},\beta} - \langle n
\rangle_\mathrm{FA,\beta}^2 = 
} \nonumber \\
&& \frac{4}{1+e^{-\beta}}\left(\langle
n_{it} n_{jt'} \rangle_{\mathrm{AA},\beta'} -\langle n
\rangle_\mathrm{AA,\beta'}^2 \right) ,
\label{equ:aa_fa_prop}
\end{eqnarray}
where $\beta$ is the inverse temperature in the FA model and $\beta'$
the inverse temperature in the AA model.  Equation
(\ref{equ:aa_fa_prop}) holds when
\begin{equation}
e^{-\beta'}=\frac{ \sqrt{1+e^{-\beta}}-1}{\sqrt{1+e^{-\beta}}+1 } .
\label{equ:aa_temp}
\end{equation}

\begin{figure}[t]
\begin{centering}
\includegraphics[width=8.5cm]{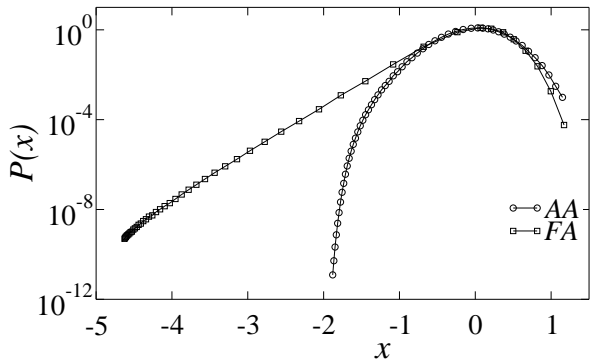}
\caption{Distribution of (reduced) box magnetization in the FA and AA 
models.
The reduced variable $x=(m-\langle m \rangle)/\sigma_m$ where
$\sigma_m^2 \equiv
\langle (m-\langle m \rangle)^2 \rangle = c(1-c)L^{-1}$
is the variance of the instantaneous magnetisation.
Parameters are $L=60$, $t_{\rm obs}=320$, $N=180$; in the FA model
$\beta=1$; in the AA model $\beta'$ is given by
(\ref{equ:aa_temp}) with $\beta=1$.  
For the AA model, $P(m)$ is close to Gaussian. 
The standard deviation $\sigma_m$ is \emph{not} trivially related to the
variance of the box magnetisation, so the fact that the Gaussian parts
of the two distributions are very similar is a non-trivial consequence
of the exact mapping between the two models.}
\label{fig:aa_fa}
\end{centering}
\end{figure}

It is clear from (\ref{equ:aa_fa_prop}) that the scaling behaviour and
critical properties of the FA and AA models are the same.  However,
their activity or magnetisation distributions are different, 
as shown in Fig.~\ref{fig:aa_fa}.  
The exponential tail is absent in the AA case.
Moreover, the persistence functions of the two models are different
[the persistence function is the probability that a site does
not flip at all in an interval of length $t$].  The fluctuations
responsible for the tails in $P(m)$ can be linked to the decoupling of
exchange and persistence times in the FA model \cite{JungGC05}, a
feature which is absent in the AA case.  
We argue that there are important differences between
the dynamics of the FA and AA models: 
subsystem observables like $P(m)$ make these
differences clear where two-point functions like $\langle n_{i\tau}
n_{j0} \rangle$ do not.

\section{Dynamical action and thermodynamic analogy}
\label{sec:action}

We have shown that the distribution of subsystem magnetisations in the
FA model has an exponential tail at small magnetisation.  We also
showed that this tail comes from rare mesoscopic regions of space-time
in which there are no relaxation events. In this section we study the
statistical mechanics of the space-time configurations of the
subsystem. We draw an analogy between these space-time statistics and
the thermodynamic statistics of a system near a phase coexistence
boundary.

\subsection{Thermodynamic analogy}

To make this analogy concrete, we define the probability of a
trajectory being generated by some (Markovian) dynamical rules:
\begin{eqnarray}
P_\mathrm{traj} &=& \prod_{\tau=1}^{Nt_{\rm obs}-1}
 W(s_{\tau+1}| \{s_{\tau}, n_{0,\tau},n_{L+1,\tau} \})
 \nonumber \\ && \times P_0(s_1) P_\mathrm{bc}(n_{0,t},n_{L+1,t}) ,
\end{eqnarray}
where $s_\tau= \{n_{1\tau}, \dots, n_{L\tau}\}$ represents
the state of the system at time $\tau$; the
transition probabilities are denoted by $W(s'|\{s,n_0,n_{L+1}\})$; 
$P_\mathrm{bc}(n_{0,t},n_{L+1,t})$ is the
probability of the trajectory of the two boundary spins; and
$P_0(s_1)$ is the probability of the initial condition ($s_1$).  We
note that the transition probabilities into a state $s_{\tau+1}$
depend on the state of the boundary spins as well as the state
$s_\tau$.

It is convenient to define the dynamical action,
\begin{equation}
\actionE_\mathrm{traj} = -\ln P_\mathrm{traj} ,
\label{equ:def_action}
\end{equation}
%
and the dynamical partition sum,
\begin{equation}
\mathcal{Z}_b = \sum_\mathrm{traj} e^{-b\actionE_\mathrm{traj}}
\label{equ:Zb}
\end{equation} 
which is equal to unity when $b=1$.
The extensivity properties of the action $\actionE$ will be discussed
below. To make the correspondence with a thermodynamic partition sum, we
simply define
\begin{equation}
Z = \sum_\mathrm{conf} e^{-\beta U_\mathrm{conf}} ,
\end{equation}
where the sum is over configurations of a system; $U_\mathrm{conf}$ is
the energy of a configuration; and $\beta$ is the inverse temperature.
The application of thermodynamic formalism to dynamical objects such as 
$\mathcal{Z}_b$ is found in the work of Ruelle \cite{RuelleBook}. 
A very recent and detailed discussion
of its application to Markov chains is given in \cite{Lecomte06}; the 
parameter $s$ of that paper is $1-b$ in our notation; 
the 
Kolmogorov-Sinai entropy $h_\mathrm{KS}$ of the Ruelle 
formalism
is the mean action, $\langle \actionE \rangle$ in our notation.

To treat phase coexistence in ordinary thermodynamics, 
one may decompose the partition function into its pure states
\cite{ParisiStatMech} 
\begin{equation}
Z = \sum_\alpha Z_\alpha
\end{equation}
where $Z_\alpha$ is the contribution of pure state $\alpha$ to the partition
sum.  For example, in the Ising system then the two pure states have
positive and negative magnetisations, and statistical
weights $Z_\uparrow$ and $Z_\downarrow$.  
The difference in free energy density
between the two states is
\begin{equation}
\mu' = \lim_{N\to\infty} N^{-1} \log( Z_\uparrow / Z_\downarrow ) 
.
\end{equation}
At phase coexistence, we have $\mu'=0$. The result is that subextensive
contributions to $\log( Z_\uparrow / Z_\downarrow )$ from
boundary conditions are sufficient to select one phase over the other.

By analogy, for dynamically heterogeneous systems, we make
an decomposition of $\mathcal{Z}$ into its reducible partitions
\cite{RitortS03}. If
there are two such partitions then we write
\begin{equation}
\mathcal{Z}_{b=1} = P_\mathrm{ic}(1) + P_\mathrm{ic}(2)
\label{equ:ic}
\end{equation}
where the two terms are the probabilities that an initial condition lies in 
one partition or the other.
We then define
\begin{equation}
\mu = \lim_{N,t_\mathrm{obs}\to\infty} (Nt_\mathrm{obs})^{-1} 
\log [ P_\mathrm{ic}(1) / P_\mathrm{ic}(2) ]
,
\label{equ:mu_ic}
\end{equation}
by analogy with the static free energy difference $\mu'$.

The probabilities $P_\mathrm{ic}(1)$ and $P_\mathrm{ic}(2)$
are independent of $t_\mathrm{obs}$, so
$\mu$ vanishes as $t_\mathrm{obs}$ is taken to $\infty$, 
and an initial condition on the system is sufficient to
select the active or inactive partition of the dynamics: see
figure~\ref{fig:bc}.  In a similar way, a boundary condition
is sufficient to select the up or down pure state in an Ising
system below the critical temperature. Recalling (\ref{equ:bubble_dist})
and (\ref{equ:ising_dist}), we observe that phase coexistence and 
reducibility lead to distributions with $\mu'=0$ and $\mu=0$
respectively. The free energy cost associated with large bubbles
(or down domains) comes from their boundaries and not from the bulk.

\subsection{Discussion of magnetisation distributions}
\label{sec:discuss_dist}

Fig.\ \ref{fig:bubble_sketch} illustrates how FA model trajectories
with low densities of relaxation events typically contain
a large compact ``bubble'' in which there are no relaxation events.
In the FA model then these large bubbles are more common
than homogeneous reductions in the density. Conversely, in the AA model,
the large bubbles are absent and the homogeneous regions of low
density determine the tail of  
the magnetisation distribution (recall figure~\ref{fig:aa_fa}).

The FA model is reducible, 
so the distribution of large bubbles 
will be that of (\ref{equ:bubble_dist}), with
$\mu=0$ [recall (\ref{equ:mu_ic})].
The derivation of the magnetisation distribution from the bubble
distribution depends on the ratio of $L$
and $t_\mathrm{obs}$.
For observation windows such that $vt_{\rm obs} < L$, 
trajectories with small $m_\mathrm{traj}$ contain
large bubbles that span the temporal extent of the observation window, as in
the lower right panel of figure~\ref{fig:bubble_sketch}. In that
case, we have
\begin{equation}
m \simeq c[1-(l/L)] .
\end{equation}
where $l$ is the spatial extent of the bubble.
The probability of the box containing a bubble of this size and shape is
\begin{equation}
P(l) \simeq \int_{t_{\rm obs}}^\infty \!\!\!\mathrm{d}\tau\, P(l,\tau)
.
\end{equation} 
Assuming that $\Lambda$ is not too small, so that $P(l,\tau)$ is sharply
peaked around $l = v \tau$, then
\begin{equation}
\frac{\mathrm{d}}{\mathrm{d}m}\ln P(m) \simeq 
(L/c)\left[ \sigma_1 + (\sigma_2/v) + 2(\mu/v)L(1-m/c) \right] ,
\end{equation}
for $m < c[1-(vt_{\rm obs}/L)]$.  Unless $\mu$ is
very small, the term proportional to $\mu$ will
dominate at large $L$, leading to a Gaussian distribution.
However, in the case of small $\mu$, the distribution is exponential in
$m$, and the gradient of this exponential tail is 
proportional to $L$ and independent of $t_{\rm obs}$, as observed in
Fig.\ \ref{fig:pm_simple}.  A similar argument holds in the opposite
regime of  $L < vt_{\rm obs}$ and $m<c[1-(L/vt_{\rm obs})]$, which
explains the exponential tails of Fig.\ \ref{fig:pm_l20}.

Pursuing the analogy with phase coexistence, 
we identify the active partition of the dynamics with a dense
phase and the inactive partition with a sparse phase.
The density of dynamical activity in the active phase is $\langle m
\rangle = c$. The typical length scale associated with bubbles of
the sparse phase is
$\sigma_1^{-1}$ which scales as $c^{-1}$.  Thus, the typical bubble
size and the typical spacing between excitations scale in the same
way as the temperature is reduced.  This scaling \cite{JMS06} is
consequence of detailed balance in the FA model, which implies that
excitations are uncorrelated at equal times: $\langle (n_{i\tau}-c)
(n_{j\tau}-c) \rangle = c(1-c) \delta_{ij}$.

In section~\ref{sec:dp}, we will
show how generalising the FA model to a system which
does not obey detailed balance leads to a situation in which the typical
bubble sizes are larger than the excitation spacing. 
        This situation is the usual one in systems at phase
        coexistence: the FA model is a special case because of the extra
        symmetry of detailed balance.        

\subsection{Distribution of the dynamical action}

In the analogy between static and dynamic partition sums, 
the dynamical action of Eq.~(\ref{equ:def_action})
corresponds to the energy of the static system.  
For the FA model, the transition probabilities are
\begin{equation}
 W( s_{\tau+1} | \{ s_\tau, n_{0\tau},
n_{L+1,\tau}\} ) = W_0 + \mathcal{P}_{s_{\tau+1},s_\tau} \frac{1}{N}
\sum_{i=1}^L W_i
\end{equation}
where the projector
$\mathcal{P}_{s,s'}$ takes the value of unity if $s$ and $s'$ differ
in the state of exactly one spin, otherwise it is equal to zero;
\begin{equation}
W_i = f_{i\tau} [(1-n_{i,\tau+1}) n_{i,\tau} +
n_{i,\tau+1} (1-n_{i\tau}) e^{-\beta} ]
\end{equation}
accounts for transitions into state $s$; and
\begin{equation}
W_0 = \delta_{s_{\tau+1},s_\tau} \left\{ 1-N^{-1}\sum_{i=1}^L \left[f_{i\tau}
(1-n_{i\tau})e^{-\beta} +n_i f_{i\tau} \right] \right\}
\end{equation}
accounts for transitions from state $s$. 
The symbol $\delta_{s,s'}$ is equal to unity if and only if states $s$
and $s'$ are identical, and the   
function $f_{i\tau}$ was defined in (\ref{equ:def_f}): its value is unity
if spin $i$ is free to flip; otherwise it is zero.  
These operators enforce the constraint that
only one spin flips on each time step.

Taking the limit of large $N$, and using (\ref{equ:def_action}), we
arrive at the action for an allowed trajectory:
\begin{eqnarray}
\actionE^{(N)}_\mathrm{traj} &=& \mathcal{N}_\mathrm{upflips} \ln (Ne^\beta)
+ \mathcal{N}_\mathrm{downflips} \ln (N) \nonumber \\ && - N^{-1}
\sum_{\tau=1}^{Nt_{\rm obs}} \sum_{i=1}^L f_{i\tau} [ (1-n_{i\tau})
e^{-\beta} + n_{i\tau}] \nonumber \\ && + (\hbox{boundary terms}) ,
\label{equ:def_SN}
\end{eqnarray}
where $\mathcal{N}_\mathrm{upflips}=\sum_{i\tau} n_{i,\tau+1}
(1-n_{i\tau})$ is the total number of flips from state $0$ to state
$1$ inside the observation box and, similarly,
$\mathcal{N}_\mathrm{downflips}=\sum_{i\tau} (1- n_{i,\tau+1})
n_{i\tau}$ is the number of flips from $1$ to $0$.  Of course,
trajectories containing transitions that are not allowed by the
dynamical rules have $P_\mathrm{traj}=0$ and their action is formally
infinite; we ignore them in what follows.  

We now discuss briefly the extensivity properties of the dynamical
action. The probability $P_\mathrm{traj}$ is associated with
the states of $L$ spins over a time period $t_\mathrm{obs}$. However,
the positions of the spin flip events in time are specified with
a resolution that depends on the system size, $N$. Thus, the
probability of any trajectory vanishes as $N\to\infty$ and
the action diverges logarithmically with $N$ as well as 
being extensive in $L$ and $t_\mathrm{obs}$. We therefore
introduce a coarse-graining timescale $\delta t$. That is, we define
\begin{equation}
\tilde{P}_\mathrm{traj} = \sum_\mathrm{traj'(\delta t)}
P_\mathrm{traj'} \simeq (N \delta t)^{\mathcal{N}_\mathrm{upflips}
+\mathcal{N}_\mathrm{downflips}} P_\mathrm{traj} ,
\end{equation}
where the sum is over trajectories with the same spin flips as the
original trajectory, but the flips may happen within an interval
$(\delta t)$ of their original times. The second, approximate,
equality indicates that as long as $(\delta t)$ is not too large then
the probability of all trajectories in the sum will be approximately
equal, and the number of such trajectories is $(N \delta
t)^{\mathcal{N}_\mathrm{flips}}$.  The result is that
\begin{eqnarray}
\actionE_\mathrm{traj}^{(\delta t)} &\equiv& -\ln \tilde P_\mathrm{traj}
\nonumber \\ &=& \mathcal{N}_\mathrm{upflips} \ln (1/e^{-\beta}\delta
t) + \mathcal{N}_\mathrm{downflips} \ln (1/\delta t) \nonumber \\ & &
+ N^{-1} \int_0^{t_{\rm obs}}\!\!\!\mathrm{d}\tau\, \sum_{i=1}^L
f_{i\tau} [ (1-n_{i\tau}) e^{-\beta} + n_{i\tau}] \nonumber \\ && +
(\hbox{boundary terms}) ,
\label{equ:def_Sdt}
\end{eqnarray}
where we have converted the sum to an integral, which is valid in the
limit of large $N$.  This definition of the action is parametrised by
the coarse-grained time scale $(\delta t)$. For fixed $\delta t$,
distributions of $\actionE_\mathrm{traj}^{(\delta t)} $ have a
peak whose position is independent of $N$ and extensive
in $L$ and $t_{\rm obs}$.  Our results do not depend
qualitatively on $\delta t$; the numerical results below are at
$\delta t=1$.  Finally, we have no prescription for
calculating the boundary terms in the action; these terms are
not extensive in $Lt_\mathrm{obs}$, so we neglect them in what follows.

\begin{figure}[t]
\begin{centering}
\includegraphics[width=7.6cm]{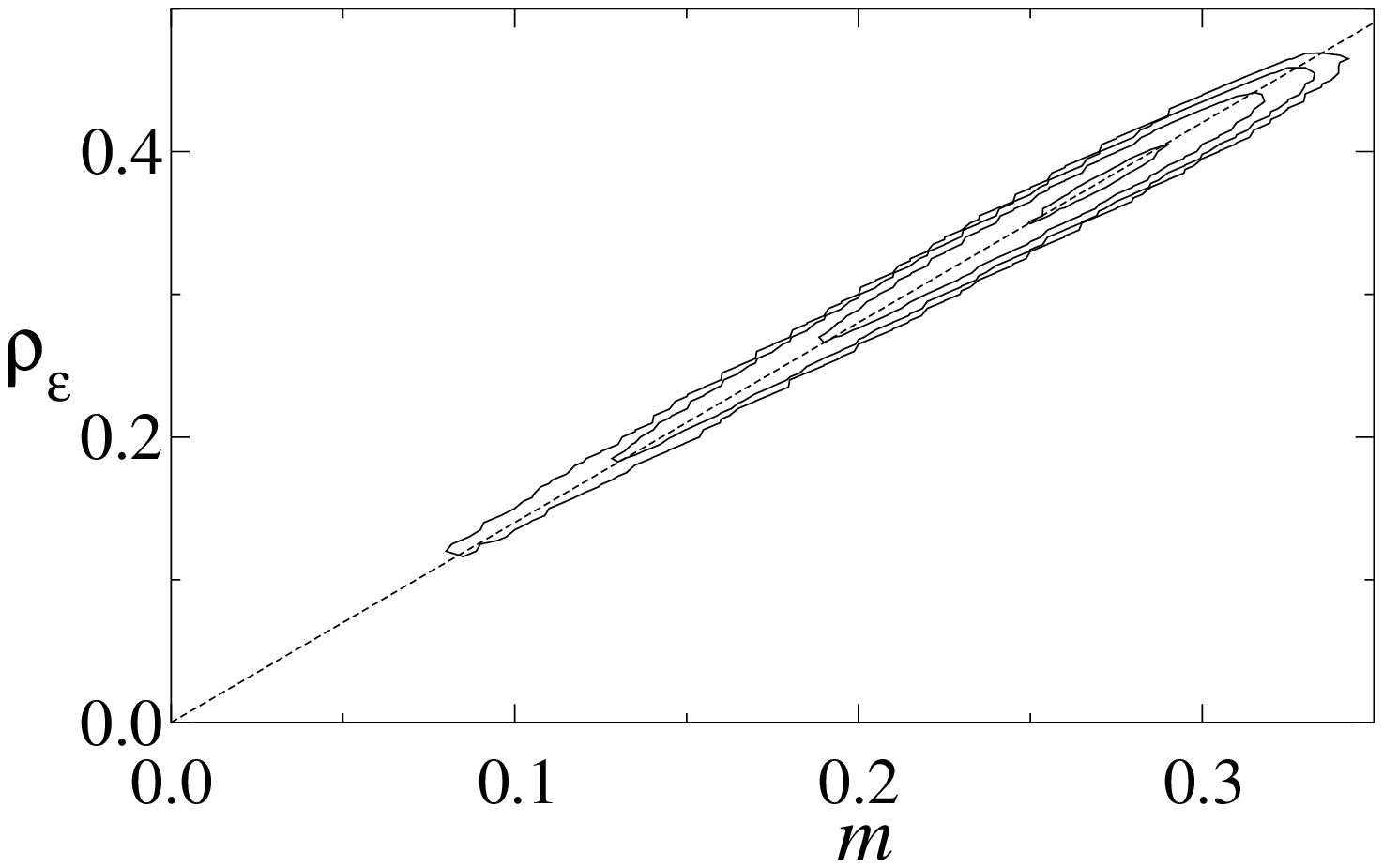}\par
\includegraphics[width=7.6cm]{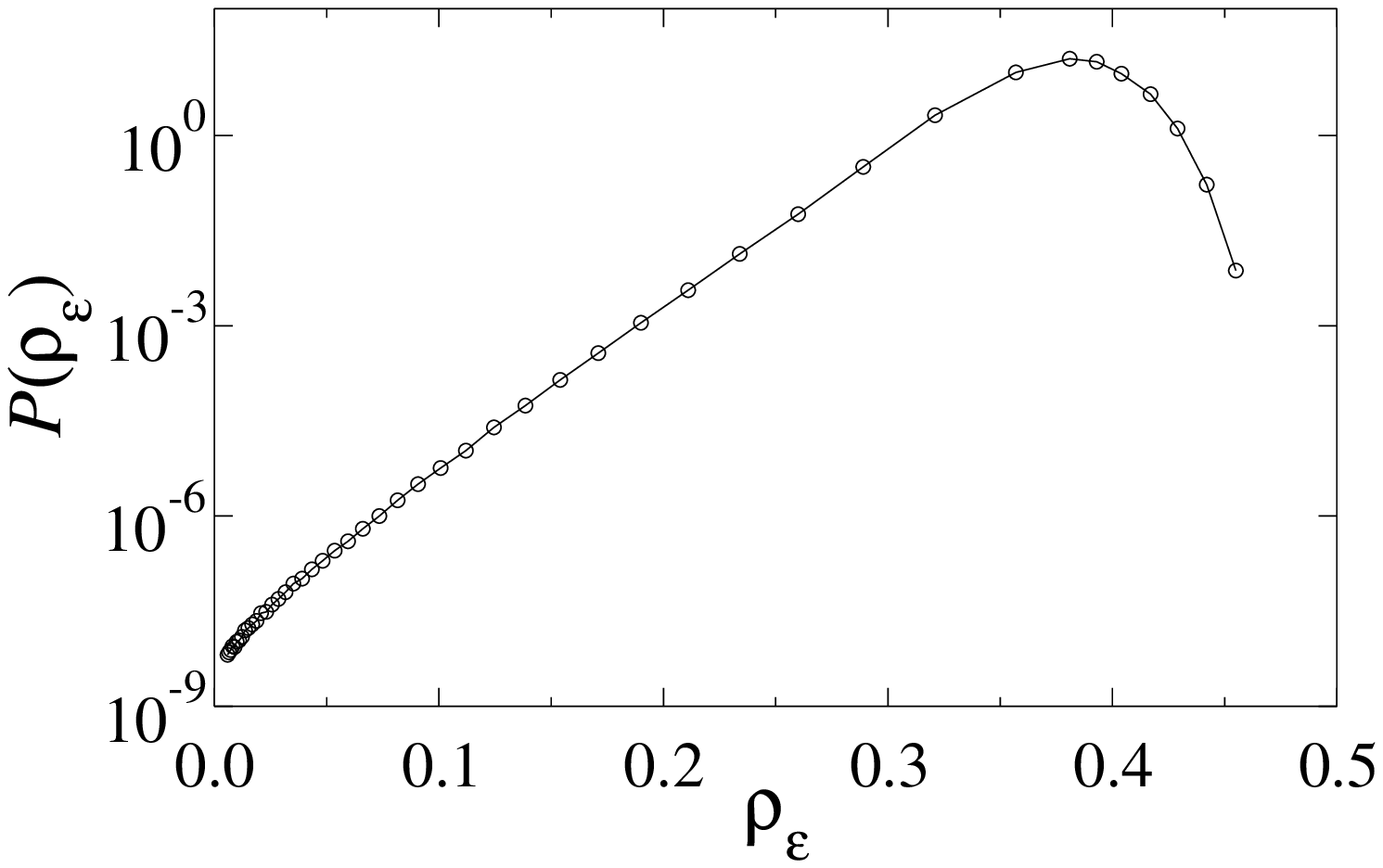}
\caption{Distribution of the action in the FA model for $L=60$,
$t_{\rm obs}=320$, obtained with $N=180$.  (Top) Contour plot of the
joint probability distribution for action density and magnetisation
$P(\rho_\actionE,m)$ (obtained from $2\times 10^7$ independent trajectories).
The contours are at $P(m,\rho_\actionE)=10^3, 10, 0.1, 2\times 10^{-3}$.  The
dotted line is the prediction (\ref{equ:sm_dist}).  (Bottom)
Distribution of the action density $P(\rho_\actionE)$ [where
$\rho_\actionE=\actionE/(Lt_{\rm obs})$].}
\label{fig:fa_action}
\end{centering}
\end{figure}

Results for the distribution of the dynamical action density,
$\rho_\actionE = \actionE^{(\delta t)}/(Lt_\mathrm{obs})$, and its
joint distribution with the box magnetisation, are shown in
figure~\ref{fig:fa_action}. We see that the distribution
of the action density is tightly correlated with that
of the magnetisation.

To understand this, we define the density of facilitated spins in
a trajectory to be
\begin{equation}
n_\mathrm{f} = (LNt_{\rm obs})^{-1}
\sum_{\tau}^{Nt_{\rm obs}} \sum_{i=1}^L f_{i\tau} .
\end{equation}

Now, facilitated down spins flip with rate $e^{-\beta}$ and facilitated
up spins flip with rate unity. Since $t_\mathrm{obs}\gg(1/c)$,
there are many such independent events within the observation
window, so that fluctuations in the number of these events
are small. Therefore,
\begin{eqnarray}
&& (LNt_{\rm obs})^{-1} \sum_{\tau}^{Nt_{\rm obs}} \sum_{i=1}^L
f_{i\tau}(1-n_{i\tau}) \simeq (1-c) n_\mathrm{f} ,
\label{equ:fac_rel1} \\
&& (LNt_{\rm obs})^{-1} \sum_{\tau}^{Nt_{\rm obs}} \sum_{i=1}^L
f_{i\tau}n_{i\tau} \simeq c n_\mathrm{f} , \\ 
&&
\mathcal{N}_\mathrm{upflips} \simeq \mathcal{N}_\mathrm{downflips}
\simeq cLt_{\rm obs} n_\mathrm{f} ,
\end{eqnarray}
where the approximate equalities indicate that the joint distribution
of each pair of observables are sharply peaked around these values
(we have verified this by measuring joint distributions
of the form shown in Fig.\ \ref{fig:fa_action}).  

Further, each up-spin is associated with two facilitated spins. Taking
into account spins that are facilitated by two separate up spins,
we have
\begin{equation}
n_\mathrm{f} \simeq m_\mathrm{traj}(2-c) ,
\label{equ:nf_rel}
\end{equation}
where the approximate equality holds in a similar sense, and 
$m_\mathrm{traj}$ is
the magnetisation or activity of the trajectory of interest, defined in 
(\ref{equ:def_mag}).

Hence, taking these results together,
we predict that the joint distribution of magnetisation
and action density is sharply peaked around
\begin{equation}
\rho_\actionE = mc(2-c) \left[ 2+\beta-2\ln (\delta t)
\right]
,
\label{equ:sm_dist}
\end{equation}
in accordance with the results of figure~\ref{fig:fa_action}.
We conclude that the exponential tail in the magnetisation
distribution is intrinsically linked with the exponential tail in the
distribution of the dynamical action. Numerical evidence for this
result was given in \cite{MerolleGC05}, but without concrete 
explanation. 

\section{Generalised model}
\label{sec:dp}

\begin{figure*}[t]
\begin{centering}
\includegraphics[width=8.0cm]{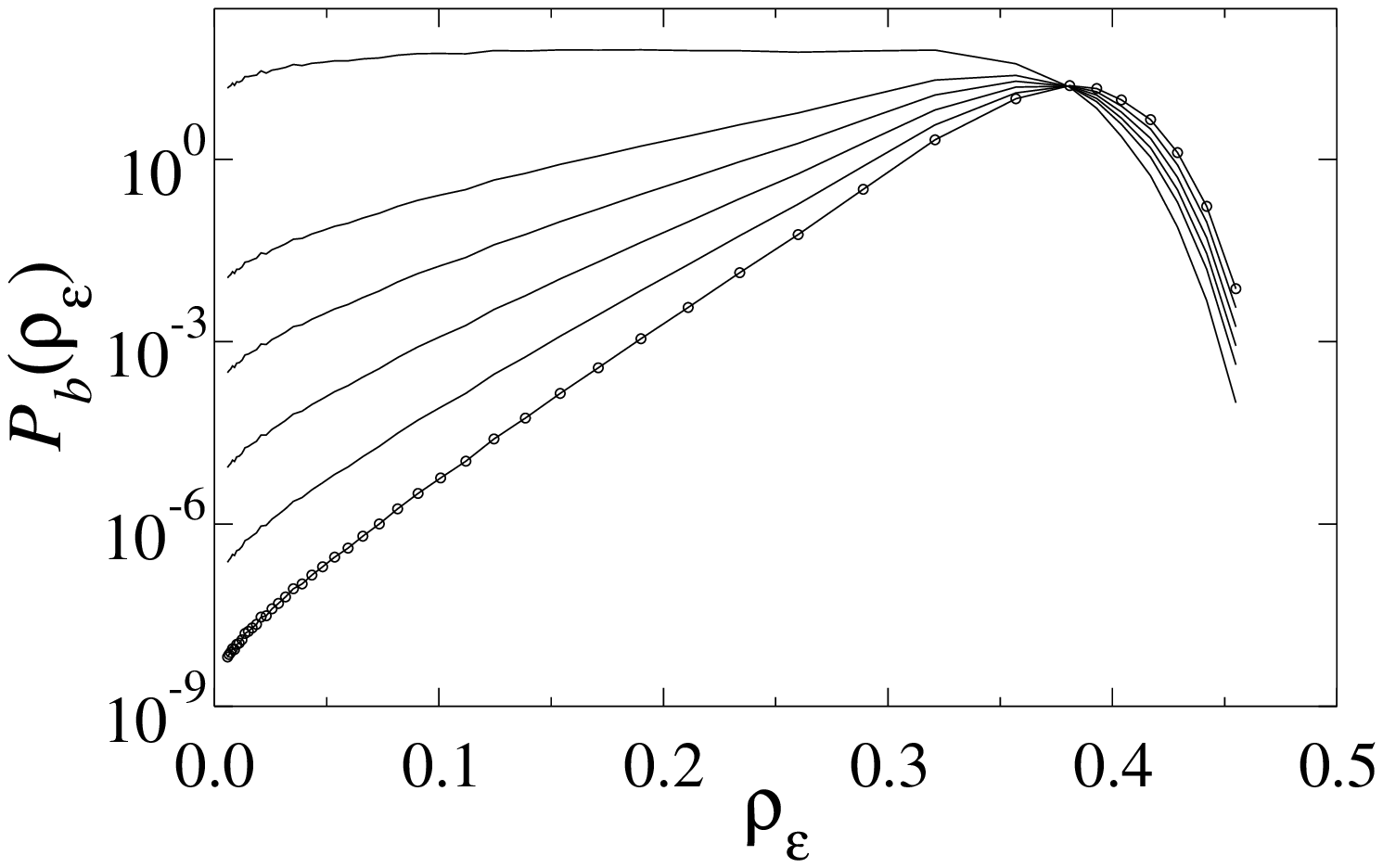}
\includegraphics[width=8.0cm]{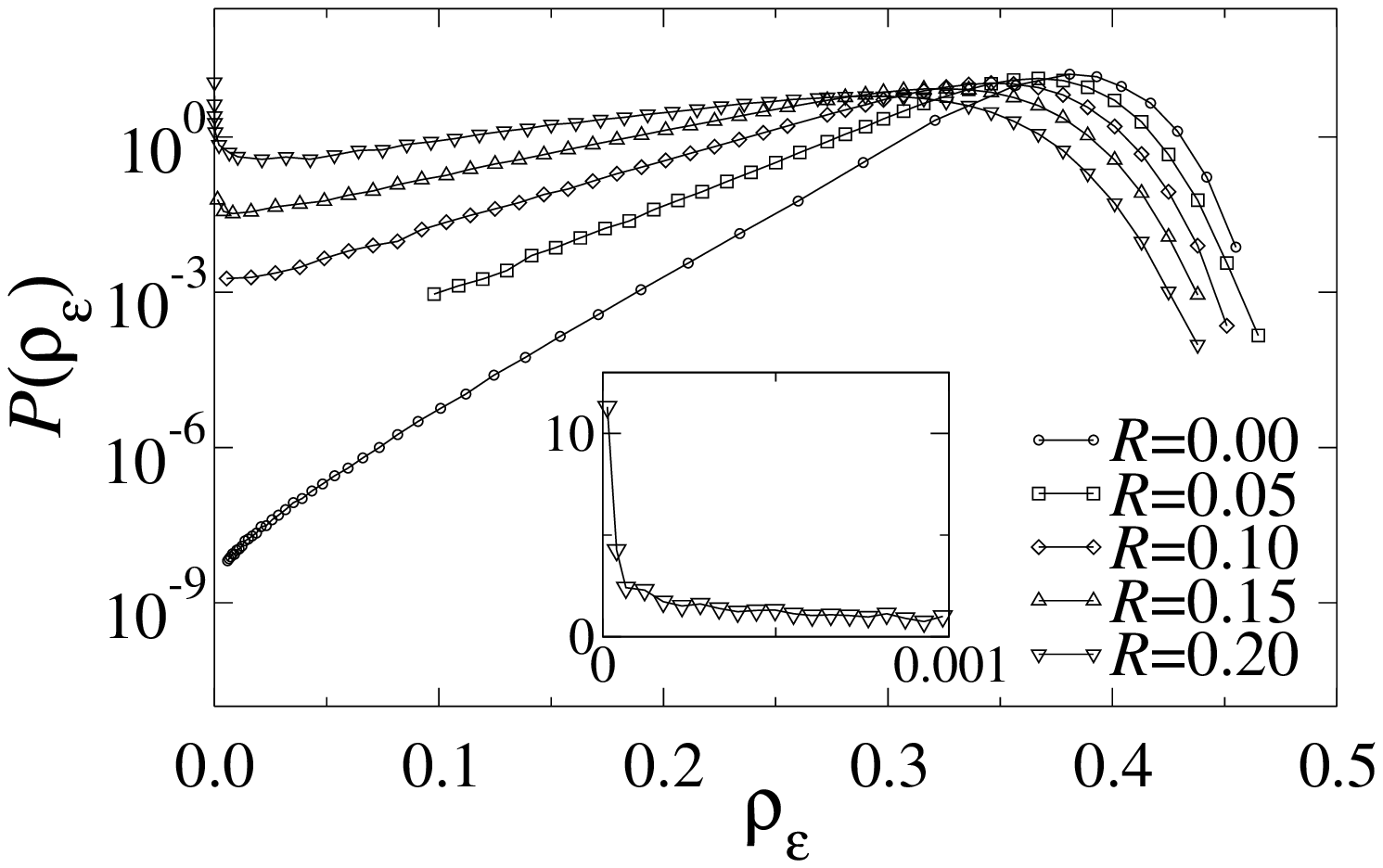}
\caption{(Left) Action distribution in the ensemble with finite
$b=(1.0,1.005,1.001,1.0015,1.002,1.003)$. The distribution at $b=1$ is
that of Fig.\ \ref{fig:fa_action} and is shown with symbols. To get
data at $b>1$ we simply use (\ref{equ:pbs}) and rescale by a constant
for convenience (this data is shown as simple lines). (Right)
Action distribution with varying $R$ at $L=60$, $t_{\rm
obs}=320$, $\beta=1$.  For $R>0$ we use $N=6L=360$ to ensure that data
is independent of $N$.  The behaviour at small $R$ is qualitatively
similar to the behaviour at small $b$ in that the gradient of the
exponential tail decreases; at larger $R$ a secondary minimum appears.
The inset shows an expanded view of the secondary minimum that is
present at $R=0.2$. Samples with the action exactly equal to zero are
omitted from the plot: the probability of this happening is of the
order of 1\% at $R=0.2$.}
\label{fig:dp_simple}
\end{centering}
\end{figure*}

The definition (\ref{equ:Zb}) motivates us to define a new ensemble for
the dynamics, in which the probabilities of trajectories are
\begin{equation}
P_{b,\mathrm{traj}} = \mathcal{Z}_b^{-1} e^{-b\actionE_\mathrm{traj}}.
\label{equ:def_b}
\end{equation}
This ensemble has the action distribution
\begin{equation}
-\ln P_b(\actionE) = - [\ln P_{b=1}(\actionE) + (b-1)\actionE] .
\label{equ:pbs}
\end{equation}
Hence, if $P_{b=1}(\actionE)$ has an exponential tail with gradient $\alpha$
then $P_{b}(\actionE)$ has a similar exponential tail with
gradient $\alpha - (b-1)$. Increasing $b$ reduces the gradient, so
large inactive regions become more common.

Observables within the $b$-ensemble can be measured 
by averaging over the dynamics at $b=1$ and reweighting the resulting
trajectories according to their action. See the left panel
of figure~\ref{fig:dp_simple}.
Since the magnetisation and action are
tightly correlated, increasing $b$ leads to a reduction in the
gradient of the exponential tail of $P(m)$.  
We associate the vanishing of this gradient with the
proliferation of trajectories with large bubbles. In \cite{MerolleGC05},
it was argued that this proliferation would appear as a first order
transition to a state with large inactive bubbles, if an appropriate
dynamics could be found to generate the $b$-ensemble.

For a given $L$ and $t_\mathrm{obs}$, a stochastic process exists
whose statistics are those of the $b$-ensemble (the probability
of each trajectory is known, and an appropriate dynamics
can be constructed).  This construction is extremely laborious,
since the transition probabilities
depend explicitly on the size of the space-time box, and on the position 
(and time) within it. Alternatively, one can simulate master equations such as
those of \cite{Lecomte06}, which do not conserve probability. Such a
simulation requires a dynamics in which `clones' of
the system are created and destroyed,
according to certain rules \cite{KurchanClone}. 

The connection to experimentally 
accessible dynamics for either alternative is not yet apparent, which 
limits analysis of the predicted first order transition.
On the other hand, 
the effects of phase coexistence can be probed experimentally by 
moving towards
the second order critical point at the end of the phase coexistence
boundary. As discussed in section~\ref{sec:discuss_dist} above,
this critical point is approached as $T\to0$ in the FA model. In that limit,
the order parameter $\langle m\rangle$
and the surface tension $\sigma_1$ vanish with the same power of $c$. The function
$P(m)$ scales according to (\ref{equ:temp_scaling}), and
it is hard to discern the two coexisting phases.

However,  
we can modify the FA model by introducing an extra process whereby sites can
flip from state $1$ to state $0$, even when unfacilitated. 
This new process breaks detailed balance, so the resulting model is not 
appropriate for an equilibrium physical system, but such models are
widely studied in a range of physical contexts \cite{Hinrichsen00}.
The rate for spins to flip from down to up is now finite at the 
critical point, and the new transition is in the directed percolation (DP)
universality class, where the effects of phase coexistence are more
apparent.

To be precise, the generalised FA model has dynamical rules
\begin{eqnarray}
n_i= 0 & \rightarrow & n_i=1 \qquad \hbox{probability } f_i e^{-\beta} ,
\nonumber
\\ n_i=1 & \rightarrow & n_i=0 \qquad \hbox{probability } f_i + r(1-f_i)
\nonumber
\end{eqnarray}
The generalised model has $r\geq0$ since all probabilities
must be positive; $r=0$ is the FA model. For $r>0$ then the system no longer
obeys detailed balance, and $\beta$ can no longer be interpreted as an
inverse temperature.  

\begin{figure}[t]
\begin{centering}
\includegraphics[width=8.5cm]{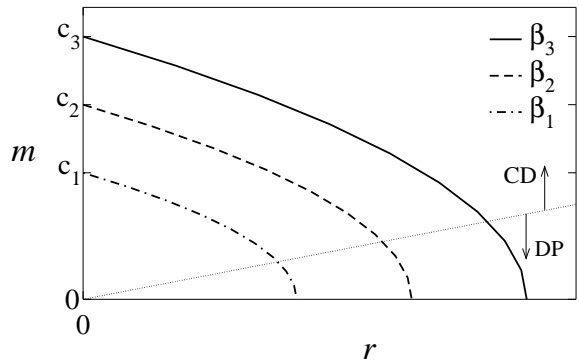}
\caption{Sketch of the steady state density in the
generalised model, as a function of $r$, for different values of $\beta$
with $\beta_1>\beta_2>\beta_3$.  The axis $r=0$ is the FA model and
the axis $m=0$ is a line of critical points. The
dotted line separates the region in which the scaling 
of directed percolation (DP) will apply from
those in which $r$ can be treated perturbatively [so the scaling will
be that of the coagulation-diffusion (CD) fixed point]. The FA model
($r=0$) is the unique case for which the critical scaling is 
coagulation-diffusion; 
for finite $r$ the relevant critical point is DP.
}
\label{fig:dp_sketch}
\end{centering}
\end{figure}

The qualitative behaviour of the model with finite $r$ is shown in
Fig.\ \ref{fig:dp_sketch}. The ``dimensionless'' parameter that
determines the effect of $r$ is the product of the death rate
and the relaxation time of the FA model:
\begin{equation}
r\tau \simeq re^{3\beta} \equiv R ,
\end{equation}
which defines $R$.  At any finite $\beta$ there is a directed
percolation transition to an active state, which occurs when
$R$ is of order unity. This transition is accompanied by a diverging
static correlation length:
\begin{equation}
\langle n_{i\tau} n_{j\tau} \rangle - \langle n_{i\tau} \rangle^2
\sim |r_i-r_j|^{z\theta-d} e^{|r_i-r_j|/\xi_\mathrm{DP}}
,
\end{equation}
where $\xi_\mathrm{DP}$ is the length scale that diverges at the
transition and $z$ and $\theta$ are DP exponents \cite{Hinrichsen00} 
(the dimensionality is $d=1$).  
On the other hand, if $R=0$ then there is a transition at
$c=0$ that is controlled by the coagulation-diffusion fixed point. In
that case we have
\begin{equation}
\langle n_{i\tau} n_{j\tau} \rangle - \langle n_{i\tau} \rangle^2
\sim \delta_{ij} .
\end{equation}
for all $c$.  We show in the figure how the crossover into the DP
critical region is always relevant if $R$ is finite, but does not
affect the behaviour if $R=0$.  We note that between two and four
dimensions a similar picture holds, but the scaling at $R=0$ will be
Gaussian \cite{JMS06}.

Turning to the dynamical action, we define
\begin{eqnarray}
\lefteqn{ \actionE^{(r)}_\mathrm{traj} = \mathcal{N}_\mathrm{upflips} \ln
(e^\beta) + \mathcal{N}_\mathrm{downflips}^{(\mathrm{death})} \ln
(1/r) } \nonumber
\\ && + \int_0^{t_{\rm obs}}\!\!\!\mathrm{d}\tau\, \sum_i
\left\{ f_{\mathrm{i\tau}} [ (1-n_{i\tau})e^{-\beta} + n_{i\tau}(1 -
r)] + n_{i\tau} r \right\} \nonumber\\
\label{equ:def_Sr}
\end{eqnarray}
by analogy with (\ref{equ:def_Sdt}),
where we have set the coarse-graining time scale $\delta t=1$ for ease
of writing. Equations~(\ref{equ:fac_rel1}-\ref{equ:nf_rel})
remain true since facilitated spins are still equilibrated at a
density close to $c=(1+e^{\beta})^{-1}$ (but note that the mean density in the system
will be less than $c$: the death process reduces the mean density).

In Fig.\ \ref{fig:dp_simple} we plot the distribution of the action
density at finite $r$ and compare it with the behaviour of the
$b$-ensemble. Since the system does not obey detailed balance the TPS
procedure becomes inefficient: data at finite $r$ is obtained by
simple binning and histogramming.  In the $b$-ensemble, the mean
action is reduced as the (finite) system crosses over from an active
to an inactive state (this crossover would be a first order phase
transition in the thermodynamic limit). On the other hand,
increasing $r$ reduces the gradient
of the exponential tail and the mean action, as the second order
phase transition is approached (at $r_c$).

\begin{figure}[t]
\begin{centering}
\includegraphics[width=4cm,height=2cm]{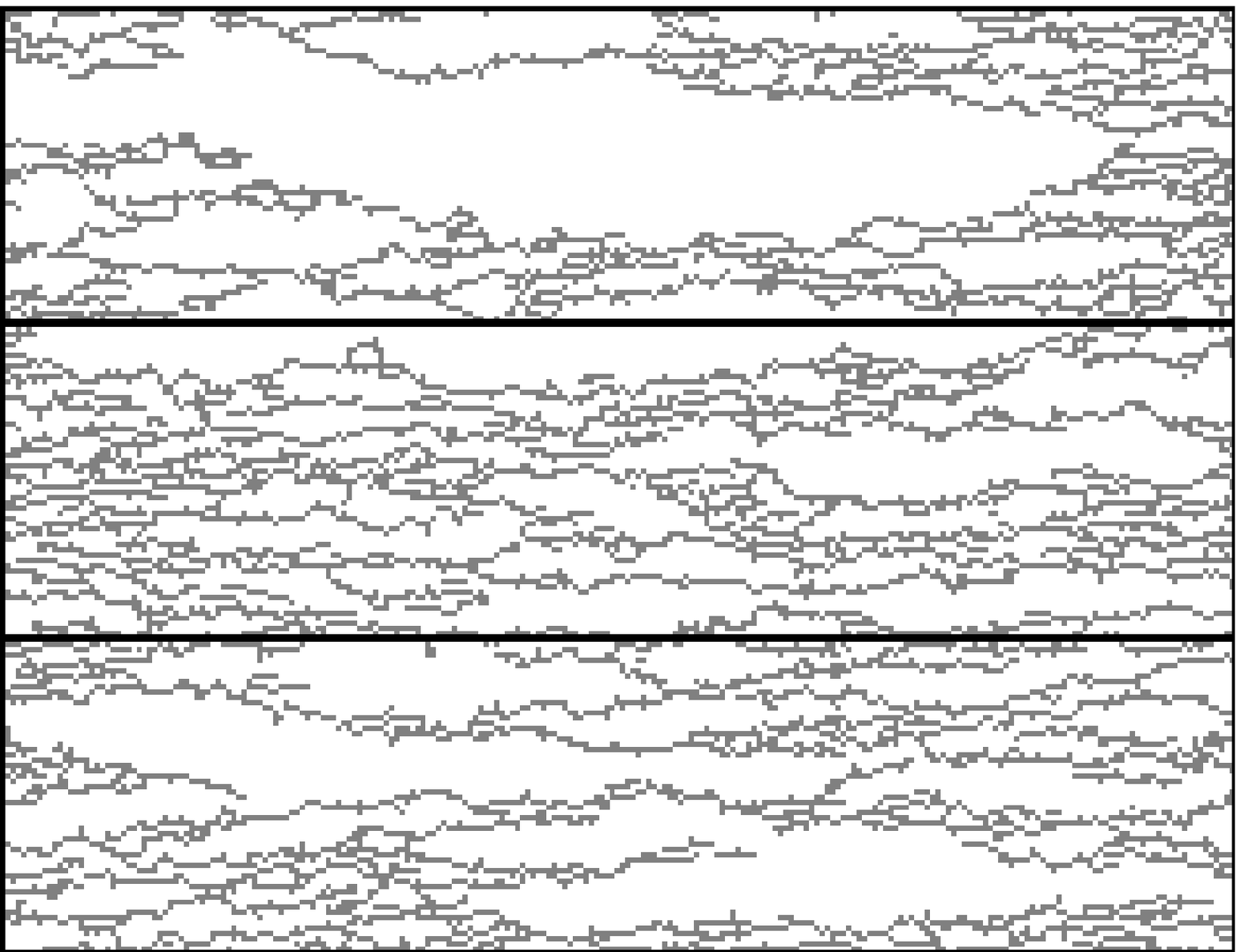}
\includegraphics[width=4cm,height=2cm]{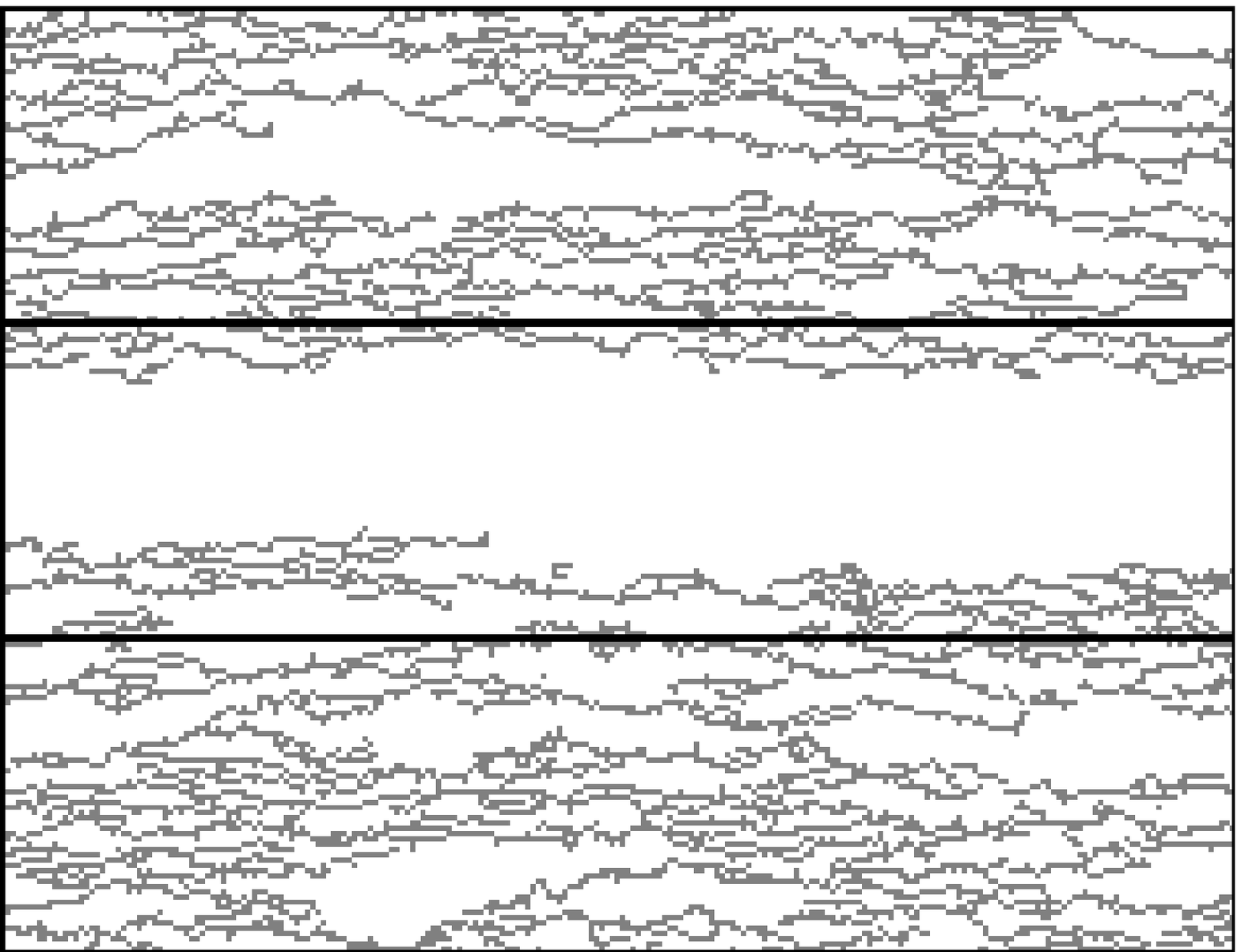}
\caption{Sample trajectories at $R=0.1$ with conditions otherwise
similar to Fig.\ \ref{fig:pm_simple} ($\beta=1$, $L=60$ $t_{\rm
obs}=320$). (Left) Sample from centre of distribution. (Right) Sample
with $m_\mathrm{traj} \sim \langle m \rangle /2$.  Clearly there are
more large inactive regions in these trajectories than in those of
Fig.\ \ref{fig:bubble_sketch}; increasing $R$ from zero leads to
proliferation of large ``bubbles''.}
\label{fig:dp_traj}
\end{centering}
\end{figure}

The key difference between the FA model and its generalisation to
finite $r$ is that the surface tension and mean magnetisation decouple
at finite $r$. On reducing the temperature in the FA model ($r=0$,
$c\to0$) we have $\sigma_1\sim\langle m\rangle\sim c$, so
the typical bubble size $\sigma_1^{-1}$ scales in the same way as the
inverse density of up spins.  In contrast, as we approach the DP
fixed point ($\beta$ finite, $r\to r_\mathrm{crit}$) then we have
\begin{equation}
\sigma_1\sim (r-r_\mathrm{crit})^{\nu_\mathrm{DP}}, \qquad \langle m
\rangle \sim (r-r_\mathrm{crit})^{\beta_\mathrm{DP}}
\end{equation}
where $(\nu_\mathrm{DP},\beta_\mathrm{DP}) \simeq (1.1,0.3)$ are the
exponents of the DP fixed point \cite{Hinrichsen00}
($\beta_\mathrm{DP}$ should not be confused with the parameter $\beta$
that enters the transition probabilities).  As criticality is approached,
the mean bubble size increases much faster than the inverse density
since $\nu_\mathrm{DP} > \beta_\mathrm{DP}$.  The up spins cluster
together, leaving large inactive regions consisting only of down
spins. Introducing the death process drives the system towards a phase
separation into active and inactive regions of space-time. This can be
seen in the trajectories of Fig.\ \ref{fig:dp_traj}.

Finally, we note that the critical DP scaling leads to a crossover
at small $R$, when the
typical size of inactive regions exceeds the observation box ($\sigma_1
L\sim 1$).  In that case, $P(\rho_\actionE)$ acquires a second peak at zero
magnetisation (see the largest value of $R$ in Fig.\
\ref{fig:dp_simple}).  This is analogous to the situation for small
boxes ($cL<1$) in the FA case (recall Fig.\ \ref{fig:pmfa_small}).  In
the thermodynamic language, it corresponds to a second minimum in the
free energy, as would be observed near a first order phase
transition. Here, the phase transition is second order, but the
finite size of the observation box means that 
the divergence of the correlation length is cut
off at the size of the observation box.

Thus, we have shown that adding a death process to the FA model strengthens
the analogy between trajectory statistics and the statistical
mechanics of phase coexistence.  
In the FA model, the only length scale
is the inverse density of up spins; if $r$ is finite then a new length
scale appears: the DP correlation length. This second length
scale allows the density of
active regions to decouple from their separation, leading to richer
behaviour than that of the FA model (which
corresponds to the particular case of the
phase coexistence in which the two length scales are equal).

Throughout this article, we have emphasised the broad applicability
of the idea of space-time phase coexistence by focussing on general
features of the models, such as microscopic reversibility and 
reducibility of the dynamics. We expect
the behaviour described here to be generic in kinetically
constrained models with reducible dynamics~\cite{RitortS03,
EisingerJ93,GarrahanC03,BerthierG05,KA-TLG}. More generally,
the extent to which this behaviour can be observed in
real structural glass-formers remains an open 
question. 

\begin{acknowledgments}
        We thank Fred van Wijland for important comments on links with 
        the Ruelle formalism.  We also benefited from discussions with
        Mauro Merolle and Tommy Miller. 
RLJ was supported in part by NSF grant No. CHE-0543158; JPG by EPSRC
grants No. GR/R83712/01 and No. GR/S65074/01, and by University of
Nottingham grant No. FEF 3024; and DC by the U. S. Department of Energy
grant No. DE-FG03-87ER13793.
\end{acknowledgments}

\begin{appendix}

\section{Transition path sampling}

We calculated probability distributions for various observables in the
FA model: these were obtained by a combination of transition path
sampling (TPS) \cite{BCD+02} and umbrella sampling \cite{FrenkelSmit}.
Here we give a brief description of the procedure used.

Transition path sampling is a Monte Carlo procedure applied to
trajectories of a dynamical system (a trajectory is a particular
realisation of the dynamics). We use it to efficiently sample
restricted ensembles of trajectories. The procedure is
well-established \cite{BCD+02}, but we present some information
regarding its application to the FA model for completeness.

Suppose that we wish to sample trajectories with magnetisation $m$
smaller than some reference value $m_1$. The simplest way to do this
is to generate statistically independent trajectories, accepting only
those that satisfy the restriction $m<m_1$. However, if $m_1$ is much
smaller than the mean of $m$ then this is inefficient.

Instead, we can generate unbiased trajectories satisfying the
restriction by deforming a set of initial trajectories, as long as the
deformations satisfy detailed balance with respect to the probability
distribution within the constrained ensemble of trajectories.  Such
deformations are called TPS moves. One possibility is to take a
trajectory of length $t_{\rm obs}$ and keep only the part with
$t<(xt_{\rm obs})$ with $0<x<1$; we the generate the rest of the
trajectory using the dynamical rules prescribed for the system (we use
a continuous time Monte Carlo algorithm \cite{NewmanBarkema}). If the
new trajectory has $m<m_1$ then it replaces the old one; otherwise the
move is rejected and we retain the old trajectory. This is a
``shooting'' move \cite{BCD+02}.  We couple this the move with the
reverse procedure in which we discard the part of the trajectory with
$t>(xt_{\rm obs})$ and regenerate the rest of the trajectory by
propagating backwards in time (since we have detailed balance then the
steady state of the FA model is invariant under time-reversal, so this
is a valid way to generate unbiased trajectories).

In addition to these moves, we also use ``shifting'' moves
\cite{BCD+02} in which we shift the trajectory in time, discarding the
parts of the new trajectory with $t<0$ or $t>t_{\rm obs}$. We then
regenerate the remaining parts of the new trajectory.  We find that
this combination of moves is quite efficient for exploring the
restricted ensembles of interest.

\section{Umbrella sampling}

We use umbrella sampling \cite{FrenkelSmit} to calculate probability
distributions by measuring ratios such as
\begin{equation} P_{i,i+1} = \frac{P(m < m_{i+1})}{P( m < m_i)}
\end{equation} 
where $m_i$ and $m_{i+1}$ are two cutoffs for the variable $m$ with
$m_{i+1}<m_i$.

Consider an ensemble of all allowed trajectories for the system, with
their statistical weights. In order to measure $P_{i,i+1}$ we sample a
restricted ensemble which contains only trajectories with $m<m_i$;
these trajectories have the same weights as they would have in the
original ensemble. We then measure the probability that $m<m_{i+1}$
within the restricted ensemble: this probability is equal to
$P_{i,i+1}$.

Our procedure is as follows:

\begin{enumerate}
\item Start with a representative set of trajectories from the
unrestricted ensemble.

\item Choose an ordered set of cutoff magnetisations
$(m_1>m_2>\dots>m_p)$ for which we will calculate the probabilities
$P_{i,i+1}$.  

\item Explore the unrestricted ensemble, measuring the fraction of
trajectories with $m<m_1$. (This is done by sampling independent
trajectories.)  

\item Once we have a good enough estimate for $P(m<m_1)$, start a
restricted ensemble with trajectories satisfying $m<m_1$. Typically we
store $N_e=100$ such trajectories.  

\item Explore the restricted ensemble using TPS, measuring the
fraction of trajectories with $m<m_2$.  Typically this takes
$N_m=100-10000$ TPS moves per ensemble member.  

\item Once we have a good enough estimate for $P_{12}$, we discard all
trajectories with $m>m_2$ and replace them by trajectories with
$m<m_2$. These trajectories are generated by continuing the TPS
procedure and accepting all trajectories that satisfy the new
constraint.  The resulting set of trajectories are not statistically
independent so we equilibrate the new ensemble by allowing it to
evolve from the biased initial condition. Typically we use around
$N_m$ TPS moves per ensemble member. We test for equilibration by
tracking the fraction of trajectories with $m<m_3$, since this will be
the quantity that we will measure on the next step.

\item We then repeat steps 5 and 6 for increasingly restricted
ensembles.  At each step, we measure $P_{i,i+1}$.
\end{enumerate}

Once we have the $P(m<m_1)$ and the set of $P_{i,i+1}$ then it is
simple to reconstruct the probability distribution of the observable
$m$.
\end{appendix}


\end{document}